\documentclass[aps,showpacs,amssymb,floatfix,pre,notitlepage]{revtex4-1}

\usepackage{mathrsfs}
\usepackage{color}
\usepackage{soul}
\usepackage[activeacute,english]{babel}
\usepackage[dvips]{graphicx}
\usepackage{amsmath,amssymb,amsfonts,calrsfs}

\definecolor{olive}{rgb}{0.5, 0.5, 0.0}

\newcommand{\sig}{\boldsymbol{\hat{\sigma}}}

\begin{document}

\title[Towards an $H$-theorem for granular gases]{Towards an
  $H$-theorem for granular gases}

\author{Mar\'ia Isabel Garc\'ia de Soria}
\author{Pablo Maynar}
\affiliation{F\'{\i}sica Te\'{o}rica, Universidad de Sevilla,
Apartado de Correos 1065, E-41080, Sevilla, Spain}
\author{St\'ephane Mischler}
\affiliation{CEREMADE, CNRS UMR 7534, Universit\'e Paris-Dauphine, 
Paris Cedex 16, F-75775 France}
\author{Cl\'ement Mouhot}
\affiliation{DPMMS, University of Cambridge,
Wilberforce Road, Cambridge CB3 0WA, UK}
\author{Thomas Rey} 
\affiliation{Laboratoire P. Painlev\'e, CNRS UMR 8524, Universit\'e Lille 1, 
59655 Villeneuve d'Ascq, France}
\author{Emmanuel Trizac}
\affiliation{LPTMS, CNRS UMR 8626, Universit\'e Paris-Sud, 91405 Orsay, France}

\begin{abstract}
  The $H$-theorem, originally derived at the level of Boltzmann
  non-linear kinetic equation for a dilute gas undergoing elastic
  collisions, strongly constrains the velocity distribution of the gas
  to evolve irreversibly towards equilibrium. As such, the theorem
  could not be generalized to account for dissipative systems: the
  conservative nature of collisions is an essential ingredient in the
  standard derivation.  For a dissipative gas of grains, we construct
  here a simple functional $\mathcal H$ related to the original $H$, that
  can be qualified as a Lyapunov functional. It is positive, and
  results backed by three independent simulation approaches (a deterministic spectral method, the
  stochastic Direct
  Simulation Monte Carlo technique, and Molecular Dynamics) indicate that it is
  also non-increasing.  Both driven and unforced cases are
  investigated.
\end{abstract}


\maketitle




\vspace*{0.75cm}

\section{Introduction}
In 1872, Boltzmann published one of his most important papers \cite{Boltzmann}.
This contribution can arguably be considered as the effective birth  
of kinetic theory, a domain pioneered by Bernoulli, Joule and Maxwell
to name but a few, and that has turned into an active field of research
in mathematics \cite{Villani02}, physics \cite{Cercignani},
and engineering \cite{Bellomo}. The interest of
reference \cite{Boltzmann} is twofold. First, 
the time evolution of the velocity
distribution function $f(\mathbf{v},t)$ for a dilute gas far from equilibrium
was derived,
under the essential assumption of molecular chaos ({\em Stosszahlansatz}),
that is, assuming the absence of correlations between the pre-collisional
velocities of colliding partners \cite{Boltzmann2,Cercignani_book}. 
For the sake of the discussion,
we omit here the spatial dependence of the velocity distribution.
Second and remarkably, 
Boltzmann also introduced
a functional $H(t) = \int f \log f \, {\rm d}\mathbf{v}$ of the instantaneous 
and time dependent distribution 
function $f$, that can be seen as a non-equilibrium entropy: the corresponding 
$H$-theorem \cite{Boltzmann2,Cercignani_book} states that this 
functional has a negative
production, and vanishes only in equilibrium \cite{rque1,rque2}. 
It therefore is a Lyapunov-like functional.
This explains in this particular context
how molecular collisions irreversibly lead to equilibrium.
Given that the underlying equations of motion are time-reversible
(with {\em elastic} collisions), this finding ignited a long lasting 
and heated debate, an account of which is not the purpose of the
present paper, and can be found in \cite{Brush,Cercignani98}
(see also \cite{Lebowitz} and references therein 
for a more technical discussion).
In essence, the statistical nature of the $H$-theorem
was not fully recognized in the early days. 
Ironically, when dissipative
collisions are considered --as e.g. is the case 
between macroscopic grains  for which the equations
of motion are irreversible \cite{g03,BP}--, 
a well defined time's arrow is present at the level of interactions,
but no $H$-like theorem could be derived so far \cite{Villani06,Bena}. 

The aim of this paper is to present strong hints that a simple
generalization of Boltzmann's original functional can be constructed,
that exhibits monotonous behavior with time and tends to zero. More
specifically, we shall be interested in the dynamical behavior of a
spatially homogeneous granular gas, made up of a collection of a large
number of grains undergoing dissipative collisions.  The grains are
treated as inelastic hard spheres, see e.g \cite{g03,barrat05} for
reviews of the rich phenomenology of this class of systems. We will
consider two cases depending on the dynamics of the grains between
collisions: the free-cooling case on the one hand, in which the grains
move freely and the stochastic thermostat case on the other hand, in
which the grains are driven by some random external force. In the first case,
it is known from particle simulations that the system reaches an
auto-similar regime in which all the time dependence in the
one-particle distribution function goes through the instantaneous
temperature (defined as the second velocity moment of the
distribution) \cite{GoSh95,MiM1,BoCe}.  In the second case, a
stationary state is reached in which the energy lost in collisions is
compensated by the energy injected by the thermostat \cite{MiM2}.

We first study an $N$-particle model where the dynamics of the
particles' velocities are treated as a Markov process. A Lyapunov
function can be identified exactly implying, under plausible
conditions, that the thermostated system reaches the stationary state
in the long-time limit. In the free-cooling case, the consequence is
that all the time dependence in the $N$-particle probability
distribution is encoded in the instantaneous temperature. This scaling
is similar to the one proposed in \cite{BDS97} at the level of the
Liouville equation and, of course, implies the one at the level of the
one-particle distribution.  Inspired by the previous model, we propose
a Lyapunov functional for the inelastic homogeneous non-linear
Boltzmann equation that describe the dynamics of the system in the low
density limit and for $N\to\infty$ \cite{BDS97,BP}. The functional is
measured in three independent and complementary numerical techniques, 
with the result that it is always a
nonincreasing function of time. These results point in the same
direction as those of \cite{mpv13}, which focussed on
thermostatted systems with emphasis on simplified collision models or
kernel.

The structure of the paper is the following. In section \ref{sec2}, the 
$N$-particle model is introduced and the corresponding $H$-functional 
identified. This is then analyzed in section \ref{sec3} at the level of 
the one-particle distribution function to in turn define, in section 
\ref{sec:conj-liap-funct}, the candidate Lyapunov function that is a central object in our 
work. In section \ref{sec:conj-liap-funct}, simulation results are also shown. 
Finally, the last section contains some summarizing remarks. 

\section{$N$-particle description}\label{sec2}

We consider a system of $N$ inelastic particles 
of mass $m$, diameter $\sigma$ and spatial dimension $d$. We will assume that 
a microstate is specified by giving the velocities of the $N$ particles, 
$\mathbf{V}(t)\equiv\{\mathbf{v}_i(t)\}_{i=1}^N$, at a given time $t$. The 
dynamics is Markovian, generated by the following rule: two particles $i$ and $j$ are 
chosen at random, together with a unit, center-to-center vector $\sig$,
from which the postcollisinal velocities
$\mathbf{v}_i', \mathbf{v}_j'$  follow
\begin{eqnarray}\label{rc1}
\mathbf{v}_i'&=&\mathbf{v}_i-\frac{1+\alpha}{2}(\sig\cdot\mathbf{v}_{ij})\sig, \\
\label{rc2}
\mathbf{v}_j'&=&\mathbf{v}_j+\frac{1+\alpha}{2}(\sig\cdot\mathbf{v}_{ij})\sig.
\end{eqnarray}
Here, $\mathbf{v}_{ij}=\mathbf{v}_i-\mathbf{v}_j$, and $\alpha$ is the 
coefficient of normal restitution that will be considered velocity-independent. 
It fulfills $0<\alpha\le 1$ and  collisions are elastic for $\alpha=1$. In 
the hard particle model, the collision frequency is
proportional to $(\sig\cdot\mathbf{v}_{ij})$. Nevertheless, we will consider a 
more general model in which this frequency is proportional to 
$(\sig\cdot\mathbf{v}_{ij})^\gamma$ where $\gamma$ is a fixed positive parameter 
($\gamma\ge 0$) \cite{ErTB06}. Then, the hard particles model is obtained for $\gamma=1$, 
while Maxwell molecules are for $\gamma=0$. Two 
cases shall be considered, depending on the dynamics of the grains between collisions.

\subsection{The unforced system}

Let us first address the free-cooling case. With the dynamics specified 
above, trajectories are generated and the state of 
the system is described in terms of the $N$-particle distribution function, 
$\rho_N(\mathbf{V},t)$. The evolution equation for this distribution is the 
generalization for inelastic collisions of the  Kac's equation \cite{Kac}
\begin{equation}\label{kacEq}
\frac{\partial}{\partial t}\rho_N(\mathbf{V},t)
=\frac{K}{N}\sum_{i<j}L(\mathbf{v}_i,\mathbf{v}_j)\rho_N(\mathbf{V},t), 
\end{equation}
where we have introduced the operator 
\begin{equation}\label{defL}
L(\mathbf{v}_i, \mathbf{v}_j)\rho_N(\mathbf{V})=\int 
(\sig\cdot\mathbf{v}_{ij})^\gamma
\left[\frac{1}{\alpha^{\gamma+1}}b_{\sig}(i,j)^{-1}-1\right]\rho_N(\mathbf{V},t)
\, {\rm d}\sig , 
\end{equation}
and $K$ is a parameter of the model with the only restriction that 
$K(\sig\cdot\mathbf{v}_{ij})^\gamma$ has dimensions of inverse of time 
(in the hard sphere case it is $K=n\sigma^{d-1}$, where $n$ is the density). 
The operator $b_{\sig}(i,j)^{-1}$ acts on any function of $\mathbf{V}$ replacing 
$\mathbf{v}_i$ and $\mathbf{v}_j$ by the precollisional velocities, i.e.
\begin{equation}
b_{\sig}(i,j)^{-1}f(\mathbf{V})=
f(\mathbf{v}_1,\dots,\mathbf{v}_{i-1},\mathbf{v}_i^*,\mathbf{v}_{i+1},\dots,
\mathbf{v}_{j-1},\mathbf{v}_j^*,\mathbf{v}_{j+1},\dots,\mathbf{v}_N),
\end{equation}
found from inverting the law \eqref{rc1}-\eqref{rc2}
\begin{eqnarray}
\mathbf{v}_i^*&=&
\mathbf{v}_i-\frac{1+\alpha}{2\alpha}(\sig\cdot\mathbf{v}_{ij})\sig, \\
\mathbf{v}_j^*&=&
\mathbf{v}_j+\frac{1+\alpha}{2\alpha}(\sig\cdot\mathbf{v}_{ij})\sig. 
\end{eqnarray}

It can be seen that Eq. (\ref{kacEq}) admits a special solution in 
which all the time dependence in the distribution function is subsumed in the 
granular temperature, $T(t)$, defined as 
\begin{equation}
\frac{d}{2}NT(t)=\int \, {\rm d}\mathbf{v}\frac{m}{2}\sum_{i=1}^Nv_i^2
\rho_N(\mathbf{V},t). 
\end{equation}
By dimensional analysis, this means that
\begin{equation}\label{sHCS}
\rho_N(\mathbf{V},t)=\frac{1}{v_H(t)^{dN}}
\varphi_N\left[\frac{\mathbf{V}}{v_H(t)}\right], 
\end{equation}
where $v_H(t)\equiv\left[\frac{2T(t)}{m}\right]^{1/2}$ is the thermal velocity. 
In Appendix \ref{a1}, a consistent equation for $\varphi_N$ is obtained and the 
equation for the temperature is analyzed. For $\gamma>0$,
the temperature behaves for long times as 
\begin{equation}\label{tAsin}
T(t)\approx\left(\frac{\gamma\bar{B}}{2}t\right)^{-2/\gamma}, 
\qquad\textrm{for $t\to \infty$}, 
\end{equation}
where $\bar{B}$ is a constant. This means that, for $\gamma>0$, the temperature 
`forgets' the initial condition in the long time limit \cite{l01,brm04}. This 
important property suggests to work in the following dimensionless variables
\begin{equation}
s=K\int_0^tdt'v_0^\gamma(t'), \qquad 
\mathbf{C}=\frac{\mathbf{V}}{v_0(t)}, 
\end{equation}
where we have introduced an ``effective'' thermal velocity (proportional at long times
to $v_H$)
\begin{equation}
v_0(t)=\left(\frac{2}{m}\right)^{1/2}
\left(\frac{\gamma\widetilde{B}}{2}t\right)^{-1/\gamma}, 
\end{equation}
with $\widetilde{B}$ an arbitrary constant. Then, the actual thermal velocity 
is proportional to the effective one in the long time limit. The evolution 
equation for the distribution function in the new variables, 
$\phi_N(\mathbf{C},s)$, 
\begin{equation}
\phi_N(\mathbf{C},s)=v_0(t)^{dN}\rho_N(\mathbf{V},t), 
\end{equation}
is
\begin{equation}\label{kacEqC}
\frac{\partial}{\partial s}\phi_N(\mathbf{C},s)
=\frac{1}{N}\sum_{i<j}L(\mathbf{c}_i,\mathbf{c}_j)\phi_N(\mathbf{C},s)
-\frac{B}{2}\frac{\partial}{\partial\mathbf{C}}\cdot\mathbf{C}
\phi_N(\mathbf{C},s), 
\end{equation}
where $B=\widetilde{B}/K$ is a dimensionless constant. 
This equation is equivalent to the one of an inelastic system (with a 
collision rule given by Eqs. (\ref{rc1})-(\ref{rc2})) whose particles 
are accelerated with a force proportional to the velocity, 
i.e. $\frac{d\mathbf{c}_i}{ds}=\frac{B}{2}\mathbf{c}_i$, and is 
usually called Gaussian thermostat. 

Let us analyze Eq. (\ref{kacEqC}) to establish the conditions under 
which a stationary state is reached in the long time limit. 
General results pertaining to 
master equations do apply to the probability 
distribution, $\phi_N(\mathbf{C},s)$, see e.g. \cite{kampen}. Let us \emph{assume} that there exists 
a stationary solution of Eq. (\ref{kacEqC}), $\phi_N^{st}(\mathbf{C})$, which 
fulfills
\begin{equation}
\frac{1}{N}\sum_{i<j}L(\mathbf{c}_i,\mathbf{c}_j)\phi_N^{st}(\mathbf{C})=
\frac{B}{2}\frac{\partial}{\partial\mathbf{C}}\cdot\mathbf{C}
\phi_N^{st}(\mathbf{C}). 
\end{equation}
Then, we consider a convex-up function $h(x)$ ($h''(x)\ge 0$), bounded from below and defined for $x\ge 0$,
from which the $\mathcal{H}_N$-functional follows\ as
\begin{equation}
\mathcal{H}_N(s)=\int d\mathbf{C}\phi_N^{st}(\mathbf{C})
h\left[\frac{\phi_N(\mathbf{C},s)}{\phi_N^{st}(\mathbf{C})}\right]. 
\end{equation}
It is shown in Appendix \ref{a2}  that 
\begin{eqnarray}
\frac{d\mathcal{H}_N(s)}{ds}=\frac{N-1}{2}\int d\mathbf{C}\int d\sig
(\mathbf{c}_{12}\cdot\sig)^\gamma\phi_N^{st}(\mathbf{C})
\left\{
h\left[\frac{\phi_N(\mathbf{C}',s)}{\phi_N^{st}(\mathbf{C}')}\right]
\right.
\nonumber\\
\left.
-h\left[\frac{\phi_N(\mathbf{C},s)}{\phi_N^{st}(\mathbf{C})}\right]
+h'\left[\frac{\phi_N(\mathbf{C}',s)}{\phi_N^{st}(\mathbf{C}')}\right]
\left[
\frac{\phi_N(\mathbf{C},s)}{\phi_N^{st}(\mathbf{C})}
-\frac{\phi_N(\mathbf{C}',s)}{\phi_N^{st}(\mathbf{C}')}
\right]
\right\}, 
\end{eqnarray}
where we have introduced the notation 
$\mathbf{C}'=\{\mathbf{c}_1'\mathbf{c}_2', \mathbf{c}_3,\dots,\mathbf{c}_N\}$, 
and use has been made of the invariance of $\phi_N(\mathbf{C},s)$ under the change of 
labels $\mathbf{c}_i\leftrightarrow\mathbf{c}_j$ for any $i$ and $j$.  
As $h(x)$ is a convex function, the integrand is negative and $\mathcal{H}_N$ 
decreases monotonically in time. On the other hand, as $\mathcal{H}_N$ is 
bounded from below, it must reach at long times a limit in which $\frac{d\mathcal{H}_N(s)}{ds}=0$. In 
this limit, the distribution is 
$\phi_N(\mathbf{C},\infty)\equiv\lim_{s\to\infty}\phi_N(\mathbf{C},s)$ that 
fulfills
\begin{equation}\label{igualdad}
\frac{\phi_N(\mathbf{C},\infty)}{\phi_N^{st}(\mathbf{C})}=
\frac{\phi_N(\mathbf{C}',\infty)}{\phi_N^{st}(\mathbf{C}')}, 
\qquad\textrm{for all $\mathbf{C}$, $\sig$}. 
\end{equation}
Note that, due to the invariance property of $\phi_N(\mathbf{C}, s)$ alluded to
above, Eq. (\ref{igualdad}) is also valid when $\mathbf{C}'$ is the 
postcollisional 
velocity vector for any  pair of particles, not necessarily the pair $1$ and $2$. 
Then, if for any $\mathbf{C}$ and $\mathbf{U}$  such 
that $C^2>U^2$ and $\sum_{i=1}^N\mathbf{c}_i=\sum_{i=1}^N\mathbf{u}_i=\mathbf{0}$, 
there exists a sequence of collisions that links $\mathbf{C}$ with 
$\mathbf{U}$, we can conclude that 
$\phi_N(\mathbf{C},\infty)=\phi_N^{st}(\mathbf{C})$.

This result is important because a stationary state in the 
$s$-variable is related to a scaling of the form given by Eq. (\ref{sHCS}) in 
the original $t$-variable. Then, if the dynamics of Eq. (\ref{kacEqC}) is such 
that, for any initial condition $\phi_N(\mathbf{C},0)$, the system reaches a 
stationary state in the long time limit, the dynamics of Eq. (\ref{kacEq}) 
will be such 
that, for any initial condition, $\rho_N(\mathbf{V},0)$, the system will reach 
the auto-similar regime with 
a scaling of the form given by Eq. (\ref{sHCS}). Moreover, for $\gamma>0$, this 
holds independently of the auxiliary parameter $B$. For $\gamma=0$ 
the situation is different, but it suffices that there exists one 
$B$ for which the stationary solution of Eq. (\ref{kacEqC}) exists. 

\subsection{The driven system}

We next treat the case in which, between collisions, 
the grains are heated by a 
stochastic force modeled by a white noise. This model is referred to as the 
stochastic thermostat model and has been extensively studied in the literature 
\cite{MoSa00,vn,pago,PLMP98}. More specifically, the jump moments of the particles' 
velocities, $B_{i,j;\beta,\gamma}$, due to the thermostat are assumed to verify 
\cite{gmt09}
\begin{equation}
\label{coefB}
B_{i,j;\beta,\gamma}\equiv\lim_{\Delta t\to 0}
\frac{\langle\Delta v_{i,\beta}\Delta v_{j,\gamma}\rangle_{noise}}{\Delta t}
=\xi_0^2\delta_{ij}\delta_{\beta \gamma}+\frac{\xi_0^2}{N-1}(\delta_{ij}-1)
\delta_{\beta\gamma},
\end{equation}
for $i,j=1,\dots,N$ and $\beta, \gamma=1,\dots,d$. We have introduced the 
notation $\Delta v_{i,\beta}\equiv v_{i,\beta}(t+\Delta t)-v_{i,\beta}(t)$, 
$v_{i,\beta}(t)$ being the $\beta$-component of the particle $i$ at time $t$. The 
parameter $\xi_0^2$ is the amplitude of the noise and 
$\langle\dots\rangle_{noise}$ denotes the average over different realizations 
of the noise. The non-diagonal terms are introduced to conserve total momentum. 
Let us remark that, as discussed in Appendix \ref{a3}, if total momentum is not 
conserved, a stationary state is not possible (indeed, the
center-of-mass velocity follows then a standard Brownian motion, see also \cite{PSD13}). 
The evolution equation for the 
$N$-particle distribution function, $\rho_N(\mathbf{V},t)$, is 
\begin{equation}
\frac{\partial}{\partial t}\rho_N(\mathbf{V},t)
=\frac{K}{N}\sum_{i<j}L(\mathbf{v}_i,\mathbf{v}_j)\rho_N(\mathbf{V},t)
+\frac{\partial}{\partial t}\rho_N(\mathbf{V},t)|_{noise}, 
\end{equation}
where the first term of 
the right hand side of the equation gives the collisional 
contribution ($L(\mathbf{v}_i,\mathbf{v}_j)$ is given by Eq. (\ref{defL})), and 
the second term 
gives the contribution of the thermostat. If the jumps due to the thermostat 
are small compared with the velocity scale in which $\rho_N(\mathbf{V},t)$ 
varies, the usual conditions to derive Fokker-Planck equations are fulfilled 
and the thermostat contribution can be approximated by \cite{kampen}
\begin{equation}
\frac{\partial}{\partial t}\rho_N(\mathbf{V},t)|_{noise}\approx
\sum_{i,j=1}^N\sum_{\beta,\gamma=1}^dB_{i,j;\beta,\gamma}
\frac{\partial}{\partial v_{i,\beta}}\frac{\partial}{\partial v_{j,\gamma}}
\rho_N(\mathbf{V},t), 
\end{equation}
with $B_{i,j;\beta,\gamma}$ given by Eq. (\ref{coefB}), independently of the 
specific probability distribution of the jumps. Then, in this limit, the 
evolution equation for $\rho_N(\mathbf{V},t)$ is
\begin{equation}\label{kacEqTS}
\frac{\partial}{\partial t}\rho_N(\mathbf{V},t)
=\frac{K}{N}\sum_{i<j}L(\mathbf{v}_i,\mathbf{v}_j)\rho_N(\mathbf{V},t)
+\xi_0^2{\mathcal T}(\mathbf{V})\rho_N(\mathbf{V},t), 
\end{equation}
where the operator $ \mathcal T$ is defined as 
\begin{equation}
{\mathcal T}(V)=
\frac{1}{2}\sum_{i=1}^N\frac{\partial^2}{\partial v_i^2}-
\frac{1}{N-1}\sum_{i<j}
\frac{\partial}{\partial\mathbf{v}_i}\cdot\frac{\partial}{\partial\mathbf{v}_j}
.  
\end{equation}

As in the free-cooling case, it is convenient to work with dimensionless 
variables. We introduce
\begin{equation}
s=Kv_s^\gamma t, \qquad \mathbf{C}=\frac{\mathbf{V}}{v_s}, \qquad
v_s=\left(\frac{\xi_0^2}{K}\right)^{\frac{1}{2+\gamma}}, 
\end{equation}
the latter having dimensions of a velocity. For the sake of readability, similar names 
as for unforced systems have been employed.
In these units, 
the evolution equation for the distribution function reads
\begin{equation}\label{kacEqCTS}
\frac{\partial}{\partial s}\phi_N(\mathbf{C},s)
=\frac{1}{N}\sum_{i<j}L(\mathbf{c}_i,\mathbf{c}_j)\phi_N(\mathbf{C},s)
+\xi^2{\mathcal T}(\mathbf{C})\phi_N(\mathbf{C},s), 
\end{equation}
where $\xi^2=\frac{\xi_0^2}{Kv_s^{\gamma+2}}$ is the dimensionless amplitude 
of the noise. 

Performing a similar analysis as in the free-cooling 
case, and under the same hypothesis, it can be shown that the function
\begin{equation}\label{def:GRE}
\mathcal{H}_N(s)=\int d\mathbf{C}\phi_N^{st}(\mathbf{C})
h\left[\frac{\phi_N(\mathbf{C},s)}{\phi_N^{st}(\mathbf{C})}\right], 
\end{equation}
decays monotonically in time and that, for any initial condition, a 
stationary distribution $\rho_N^{st}(\mathbf{V})$ is reached in the long time 
limit. 


\section{One-particle description}\label{sec3}
In the previous section, we provided a description at the $N$-particle level. 
Here, we consider the $N\to\infty$ limit case, where the problem
is expected to admit a closed description in terms of the velocity distribution 
function, $f(\mathbf{c},s)$,  
\begin{equation}
\phi_{N,1} (\mathbf{c}_1,s)  \approx f(\mathbf{c}_1,s)  \quad\text{for}\quad N\gg 1, 
\end{equation}
that we take normalized to unity, where the $j$-th marginal of $\phi_N$ is defined by
\begin{equation}
\phi_{N,j} (\mathbf{c}_1, \dots , \mathbf{c}_j)  
\equiv\int d\mathbf{c}_{j+1}\dots d\mathbf{c}_N
\phi_N(\mathbf{C}).
\end{equation}
Integrating Eqs. (\ref{kacEqC}) and (\ref{kacEqCTS}) over 
$\mathbf{c}_2, \dots, \mathbf{c}_N$ and assuming the chaos property 
\begin{equation}\label{factorization2}
\phi_{N,2} (\mathbf{c}_1, \mathbf{c}_2,s) \approx 
f(\mathbf{c}_1, s)f(\mathbf{c}_2, s), \quad\text{for}\quad N\gg 1, 
\end{equation}   
the homogeneous Boltzmann equation for the two cases is obtained
(resp. in the unforced and driven cases)
\begin{eqnarray}\label{BoltzEq}
\frac{\partial}{\partial s}f(\mathbf{c}_1,s)
&=&\int d\mathbf{c}_2L(\mathbf{c}_1,\mathbf{c}_2)f(\mathbf{c}_1,s)f(\mathbf{c}_2,s)
-\frac{B}{2}\frac{\partial}{\partial\mathbf{c}_1}\cdot\mathbf{c}_1
f(\mathbf{c}_1,s), \\
\label{BoltzEqTS}
\frac{\partial}{\partial s}f(\mathbf{c}_1,s)
&=&\int d\mathbf{c}_2L(\mathbf{c}_1,\mathbf{c}_2)f(\mathbf{c}_1,s)f(\mathbf{c}_2,s)
+\frac{\xi^2}{2}\frac{\partial^2}{\partial\mathbf{c}_1^2}f(\mathbf{c}_1,s). 
\end{eqnarray} 
It is worth emphasizing that the above ``chaos" notion has been
introduced by Kac in order to formalize the idea of asymptotic
independence of particles in the limit $N \to \infty$.
 
Until now, no Lyapunov functional for Eqs. (\ref{BoltzEq}) and
(\ref{BoltzEqTS}) has been identified. Nevertheless, the analysis made
in the previous section at the $N$-particle level suggests the
following. Consider a specific example of $\mathcal{H}_N$ discussed
above. Taking $h(x)=x\log{x}$, which is bounded from below by
$ -e^{-1}$ for $x\ge 0$, we get
\begin{equation}\label{HNlog}
\mathcal{H}_N(s)=\int d\mathbf{C}\phi_N(\mathbf{C},s)
\log{\left[\frac{\phi_N(\mathbf{C},s)}{\phi_N^{st}(\mathbf{C})}\right]}. 
\end{equation}
In addition, if we assume that the $N$-particle distribution factorizes  for all times in 
terms of the one-particle probability distribution, i.e. 
\begin{equation}\label{factorizationN}
\phi_N(\mathbf{C},s)=f(\mathbf{c}_1,s)\dots f(\mathbf{c}_N,s)
\end{equation}
and
\begin{equation}\label{factorizationNphi}
\phi_N^{st}(\mathbf{C})=f^{st}(\mathbf{c}_1)\dots f^{st}(\mathbf{c}_N), 
\end{equation}
where $f^{st}$ is the stationary solution of the Boltzmann equation,
Eq.  (\ref{BoltzEq}) or (\ref{BoltzEqTS}), the functional
$\mathcal{H}_N$ becomes extensive and transforms into a functional of
the one-particle distribution function $f(\mathbf{c},s)$,
\begin{equation}  
\frac{1}{N}Â \mathcal{H}_N(s) = \int d\mathbf{c}f(\mathbf{c},s)
\log{\left[\frac{f(\mathbf{c},s)}{f^{st}(\mathbf{c})}\right]} .
\end{equation}

Let us remark that the factorization form given by
Eq. (\ref{factorizationN}) and Eq.  (\ref{factorizationNphi}) should
be understood in the sense of Eq. (\ref{factorization2}): it can
represent a good approximation at least in the $N\to\infty$ limit.  In
the elastic limit, the above argument can be completely justified and
it has been shown that
\begin{equation} \label{Hchaos}
\lim_{N\to\infty} \frac{1}{N} \mathcal{H}_N(s)  =\int d\mathbf{c}f(\mathbf{c},s)
\log{\left[\frac{f(\mathbf{c},s)}{f^{st}(\mathbf{c})}\right]},
\end{equation}
a property called as ``entropic chaos". The above limit has been
first established for a particular class of well-prepared time
independent sequences of $N$-particle densities $\phi_N$ by Kac in
\cite{Kac} (see also \cite{CCLLV}) and more recently for any sequence
of solutions to the elastic Kac's equation (\ref{kacEq}) in \cite{MiM}
(see also \cite{HauMi,Carra}). The most important difficulty in
establishing \eqref{Hchaos} lies in the proof of the convergence
\begin{equation}
\frac1N \int d\mathbf{C} f(\mathbf{c}_1) \dots  f(\mathbf{c}_N) \log \phi_N^{st}(\mathbf{C}) 
\ \to \  
\int d\mathbf{c}_1 f(\mathbf{c}_1) \log f^{st}(\mathbf{c}_1)
\end{equation}
that one can deduce from a careful use of an accurate version of the
central limit theorem.  It is worth mentioning that the limit
(\ref{Hchaos}) is not the only possible scenario. Consider for
instance the $N$-particle McKean-Vlasov model
\begin{equation}\label{McKEq}
  \frac{\partial}{\partial t}\phi_N(\mathbf{C},t)
  = \sum_{i=1}^N \hbox{div}_{\mathbf{c}_i} (A_{N,i} (\mathbf{C)} \phi_N(\mathbf{C},t)) +  \sum_{i=1}^N \Delta_{\mathbf{c}_i}   \phi_N(\mathbf{C},t)
\end{equation}
where the force field term $A_{N,i}$ is obtained from the $N$-body
Hamiltonian function $W_N$ and $2$-body Hamiltonian function $a$ by
\begin{equation}
A_{N,i} (\mathbf{C)}= \nabla_{\mathbf{c}_i} W_N(\mathbf{C)}, 
\quad W_N(\mathbf{C)} := \frac{1}{N} \sum_{i,j = 1}^N a(\mathbf{c}_i-\mathbf{c}_j) + \sum_{i= 1}^N | \mathbf{c}_i|^2.
\end{equation}
It is clear and well-known that the only positive and normalized
stationary state is the Gibbs probability measure $\phi_N^{st}$ given
by
\begin{equation}
\phi_N^{st}Â (\mathbf{C}) := \frac{1}{Z_N} \, e^{- W_N(\mathbf{C)}}, \quad Z_N := \int d\mathbf{C}  e^{- W_N(\mathbf{C)}}. 
\end{equation}
The relative entropy $\mathcal{H}_N(s)$ defined in (\ref{HNlog}) is still
a Lyapunov functional but now, under some smoothness and boundedness
assumptions on the $2$-body Hamiltonian function $a$, one can show
that the rescaled relative entropy $\mathcal{H}_N(s)$ converges to the
free energy, namely
\begin{equation}
\frac1N \mathcal{H}_N(s) \xrightarrow[N \to \infty]{} {\mathcal F}(s) := \int d\mathbf{c}f(\mathbf{c},s)
\log{\left[\frac{f(\mathbf{c},s)}{Z^{-1} \, e^{-|\mathbf{c}|^2}}\right]}
 + \frac{1}{2} \int d\mathbf{c}_1d\mathbf{c}_2f(\mathbf{c}_1,s)f(\mathbf{c}_2,s)
\, a(\mathbf{c}_1-\mathbf{c}_2). 
\end{equation}
This convergence can be rigorously justified mathematically by (1) using
the techniques of \cite{MiMW11} to prove the propagation of chaos on
any $k$-marginal for this many-particle system, (2) using the
technique in \cite[Section~7]{MiM} to prove the convergence of the
rescaled relative entropy. 
We are interested here in a dilute system -- a proviso necessary for the validity of
the Boltzmann description --, in which the precise form of the law of interaction 
between the particles is irrelevant, beyond the fact that collisions 
are dissipative. We do not expect a `Hamiltonian fingerprint' in the limit
$\lim_{N\to \infty}\mathcal{H}_N/N$, and we are then led in the next section to conjecture that
the relevant form is \eqref{Hchaos}.

Note finally the ``gap'' between the $N$-particle evolution, where
the evolution is linear and infinitely many $h$ are admissible in the
definition \eqref{def:GRE}, and the mean field limit $N =\infty$ where
it is crucial to use the extensivity of the logarithm function
(imposing $h(z) = z \ln z$) so that the relative entropy scales like
the number of particles and the rescaled relative entropy can converge
to an effective relative entropy for the limit nonlinear Boltzmann
equation.

\section{The conjectured Lyapunov function}
\label{sec:conj-liap-funct}

We are then in a position to define our Lyapunov-candidate functional
as the Kullback-Leibler distance (also called relative entropy)
\cite{CT} between the time dependent velocity distribution,
$f(\mathbf{c},s)$, and its long time limit, $f^{st}(\mathbf{c})$,
\begin{equation}
{\mathcal H}(t) = \int d\mathbf{c} f(\mathbf{c},t) 
\log\left[\frac{f(\mathbf{c},t)}{f^{st}(\mathbf{c})}
\right]. 
\label{eq:KL}
\end{equation}
A convexity argument shows that this quantity is positive \cite{CT},
and by construction, it is expected to vanish at long times. Our
central conjecture is that it does so monotonously in time, {\it i.e.}
$d{\mathcal H}/dt <0$.
It is also important to emphasize here
that with elastic collisions for which the velocity distribution
thermalizes and evolves towards
$f^{st}(\mathbf{c}) \propto \exp(-c^2)$, the above distance
${\mathcal H}(t)$ reduces to the original Boltzmann $H(t)$ functional
alluded to above, up to an irrelevant constant.  It should also be
emphasized that Eq. \eqref{eq:KL} is invariant under change of
variable $\mathbf{c} \to \psi(\mathbf{c})$ where $\psi$ is some
invertible function, an important requirement for an entropy-like
functional \cite{MT11}. 

We first sketch a heuristic argument and then perform numerical
simulations. The goal is to prove that the relative entropy production
is non-negative
\begin{align}
  \mathcal{D}_\alpha(f) := - \int d{\bf c_1} d{\bf c_2} L_\alpha({\bf c_1},{\bf
  c_2}) f({\bf c_1}) f({\bf c_2}) - (1-\alpha) \int d{\bf c_1}
  \frac{\partial^2}{\partial {\bf c_1}^2} f({\bf c_1}) \ge 0
\end{align}
where we keep track of the inelasticity in $L_\alpha$, 
defined as above. Then, assuming that evolution
nonlinear equation propagates strong regularity and decay and arguing
without full rigor on the functional spaces level, we search for
minimisers critical points $f^*$ of $\mathcal{D}_\alpha$ with strong
regularity and decay. We then have heuristically 
\begin{align}
  \mathcal{D}_\alpha(f^*) \sim \mathcal{D}_1(f^*) + O(1-\alpha)
\end{align}
for $\alpha$ close to $1$, and due to entropy - entropy production
estimates, we deduce that
$f^* = f^{st}_1 +O(1-\alpha) = f^{st}_\alpha + O(1-\alpha)$, see the
argument in \cite{MiM1}. Finally by studying the Euler-Lagrange
equation satisfied by the minimisers $f^*$ and performing a
perturbative argument in the $O(1-\alpha)$ neighborhood, we prove that
$f^* = f^{st}_{\alpha}$ is the unique minimizer locally. 
The previous argument is perturbative in dissipation. Numerical data suggest, however, that 
the monotonicity of ${\mathcal H}$ goes beyond the quasi-elastic range.


We have implemented three complementary and independent simulation 
techniques to assess and illustrate our central statement that
$d{\mathcal H}/dt<0$: a spectral approach, the Direct Simulation Monte Carlo 
(DSMC) technique and Molecular Dynamics (MD) simulations. 
We now discuss each
method in more detail.
\begin{itemize}
\item  In the spectral method the non-linear Boltzmann 
equation \eqref{BoltzEq} is directly solved. The velocity distribution is truncated,
Fourier transformed (assuming periodic boundary conditions), and the evolution of each Fourier
mode $\widehat{f}_k$ is subsequently computed from e.g. a Runge-Kutta scheme.
In the driven case, the evolution is given by
\[
  \widehat{f}_k'(t) = \sum_{\substack{l+m = k \\ l,m = -M}}^M \widehat{f}_l(t) \widehat{f}_m(t) \left (\widehat{\beta}_{l,m} - \widehat{\beta}_{m,m}\right ) - C k^2 \widehat{f}_k(t),
\]
where the so-called kernel modes $\widehat{\beta}_{l,m}$ depend only on $\gamma$ and $\alpha$,
and can be precomputed and stored before solving the equation, $C$ being a nonnegative constant.
This method was first derived in \cite{PareschiPerthame96} for the elastic case,
and then extended to the inelastic case in \cite{FiPaTo05}.
It is deterministic and spectraly accurate by nature, preserves mass exactly, momentum 
and temperatures spectraly and costs $ O(M^2)$ operations.
It is moreover valid for any values of $\alpha \in (0,1]$. 
\item The DSMC method is widely used in the present context, in aeronautics,
and in microfluidics \cite{Bird}. $N$ particles follow a Kac's walk in velocity
space and in the limit of large $N$, the corresponding first 
marginal, $f(\mathbf{c},s)$, evolves according to the Boltzmann equation 
\cite{MiMW11}. The method is Monte Carlo in spirit, and thus of stochastic nature.
\item In the MD simulations the exact equations of motion 
are integrated, starting from a
given initial configuration of $N$ grains in a finite simulation
box of volume, $V$, with periodic boundary conditions \cite{AT}. This method 
does not rely on the putative validity of a kinetic description 
and by comparing to the outcome of DSMC, 
provides a stringent test of the theory and predictions under scrutiny. 
In particular, the spatial dependence is fully accounted for within MD 
--unlike in the DSMC approach used where spatial homogeneity is 
enforced from the outset-- and does not rely on the molecular chaos assumption. 
If $N\to\infty$ and in the low-density limit (or more precisely, in the Grad's 
limit) the first marginal is expected to fulfill the Boltzmann equation. 
\end{itemize}

In the simulations, the evolution of the one-particle distribution function 
has been measured for the two models, i.e. the Gaussian and stochastic 
thermostats, using  different values of the inelasticity and starting with 
different initial velocity distributions. With that, the functional ${\mathcal H}$ 
can be computed through Eq. (\ref{eq:KL}), where the knowledge of the late time distribution
$f^{st}$ 
is required. Hence, ${\mathcal H}$ cannot be obtained ``on the flight'',
but is computed after $f^{st}$ has been measured in the simulations. 
We have taken the grain's mass, $m$, as the unit of mass and the initial 
temperature, $T(0)$, as the unit of temperature. In the MD simulations the 
unit of length is the diameter of the particle, $\sigma$. We 
always considered a two-dimensional system  of $N=1000$ disks. 
The spectral method is used in $2$ dimensions of the velocity space, with $64$ modes in each space directions. 
It is known that such a number of modes gives a very good accuracy, thanks to the spectral convergence of the method.
The Gaussian thermostat case has been studied by DSMC and MD, while 
the stochastic thermostat has been addressed via DSMC and spectral methods \cite{rque55}.

\begin{figure}
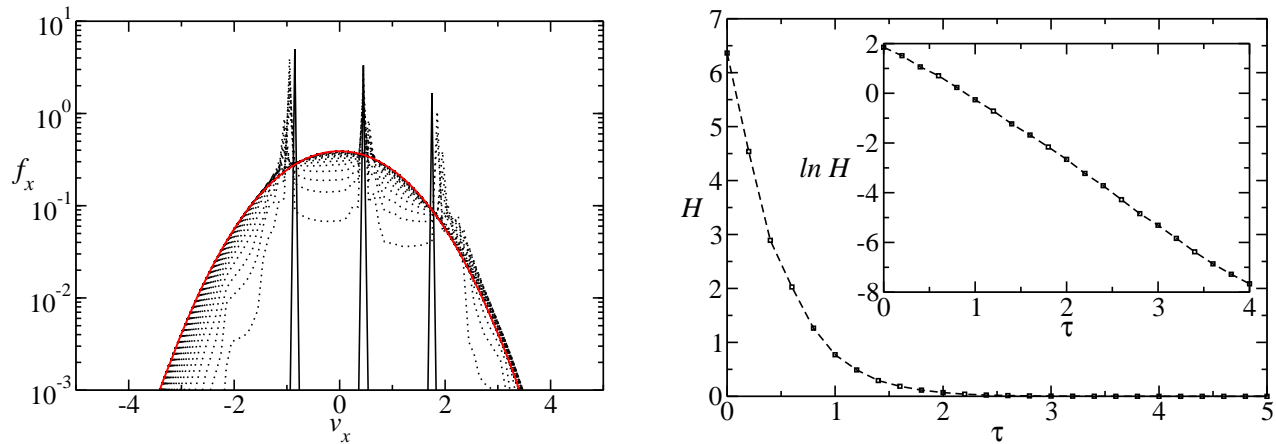

\begin{minipage}{0.49\linewidth}
\begin{center}
\includegraphics[angle=0,width=0.9\linewidth,clip]{fvx_C_I_asim_mod.eps}
\end{center}
\end{minipage}
\begin{minipage}{0.49\linewidth}
\begin{center}
\includegraphics[angle=0,width=0.9\linewidth,clip]{decaimiento_C_I_asim_mod.eps}
\end{center}
\end{minipage}
\caption{ Free cooling. Left) (color online) DSMC results for $f_x$ with an initial 
asymmetric velocity distribution made up of three sharp peaks. The parameters 
are $\alpha=0.9$ and $N=1000$ and the results have been averaged over $10^5$ 
realizations. The distribution is plotted for different values of the number 
of collisions per particle $\tau$.  The black solid line corresponds to the 
initial distribution and the bell-shaped red solid line to the distribution at 
the end of the simulation ($\tau\simeq 14$). Note that the $y$-scale is 
logarithmic to better probe the low probability region. Right)  Corresponding 
evolution of ${\mathcal H}(t)$. The inset 
shows the same data on a linear-log scale.}\label{fig1}
\end{figure}

Fig. \ref{fig1} 
displays DSMC results for a system with dissipation parameter $\alpha=0.90$ 
heated by the Gaussian thermostat with $B$ chosen to have unit stationary 
temperature. The results have been averaged over $10^5$ realizations and the 
initial distribution has been taken asymmetric with three peaks:
\begin{equation}
\label{eqAssym}
f(\mathbf{v},0)=\frac{3}{6}\delta(\mathbf{v}-\mathbf{u}_1)
+\frac{2}{6}\delta(\mathbf{v}-\mathbf{u}_2)
+\frac{1}{6}\delta(\mathbf{v}-\mathbf{u}_3), 
\end{equation}
with $\mathbf{u}_2=\frac{1}{3}\left[\frac{T(0)}{m}\right]^{1/2}(1,1)$, 
$\mathbf{u}_1=-3\mathbf{u}_2$, and $\mathbf{u}_3=5\mathbf{u}_2$. In the left 
side of the figure, the distribution function for $v_x$, defined as 
$f_{x}(\mathbf{v}_x,t)=\int \, {\rm d}{v}_y f(\mathbf{v},t)$, has 
been plotted for different values of the number of collisions per particle, 
$\tau$. 
Clearly, the behavior of 
${\mathcal H}(t)$ on the right hand side is compatible with an asymptotic vanishing for $t\to\infty$, 
which simply indicates that $f$ tends towards $f^{st}$. 
More interestingly, ${\mathcal H}$ is non increasing, from the shortest times, to
the largest ones one can reach in the simulations. 

\begin{figure}
\begin{minipage}{0.7\linewidth}
\begin{center}
\includegraphics[angle=0,width=0.9\linewidth,clip]{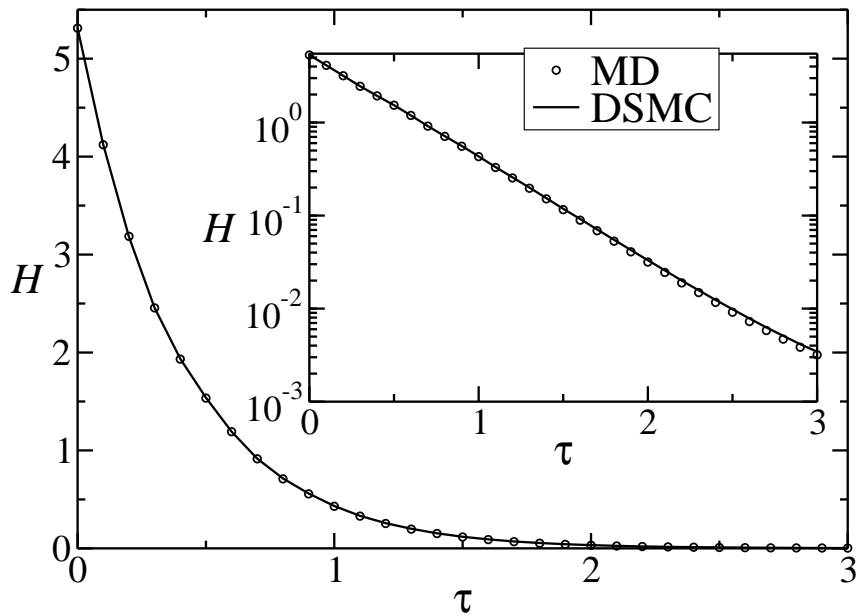}
\end{center}
\end{minipage}
\caption{Free cooling. Evolution of ${\mathcal H}(t)$ as a function of the number of collisions 
per particle for MD and DSMC simulations, for an asymmetric initial velocity 
distribution.  
Here $\alpha=0.8$, $N=1000$ and the results have been averaged over 
$5 \times 10^3$ realizations. The inset shows the same data on a linear-log 
scale, to emphasize the long time behavior.}\label{fig2}
\end{figure}

In Fig. \ref{fig2}, a comparison between MD and DSMC results is shown for 
a system with $\alpha=0.80$. The initial distribution is the same asymmetric 
distribution as in the previous case and, again, $B$ has been chosen to 
have unit stationary temperature. The results have been averaged over 
$5 \times 10^3$ realizations in the two kinds of simulations. 
The density in the MD simulations is $n=0.005\sigma^{-2}$, which corresponds to 
a rather dilute system.
The excellent agreement between MD and DSMC is important, 
not only because it again points to the monotonicity of ${\mathcal H}(t)$
but also because the MD algorithm provides a reference benchmark
(``true dynamics''), which does not rely on the hypothesis leading to
the Boltzmann equation, and in particular does not {\it a priori} assume 
the system to be homogeneous. It should be mentioned though that the
parameters chosen for MD are such that the system remains in a spatially
homogeneous state for all times (see e.g. the discussion in Refs
\cite{HCS_unstable} for the free cooling regime). Let us mention that we have 
observed the same qualitative features for a large gamut of initial 
conditions (symmetric around the velocity origin or asymmetric) and different 
values of the inelasticity in the whole range, $0<\alpha<1$. 

\begin{figure}
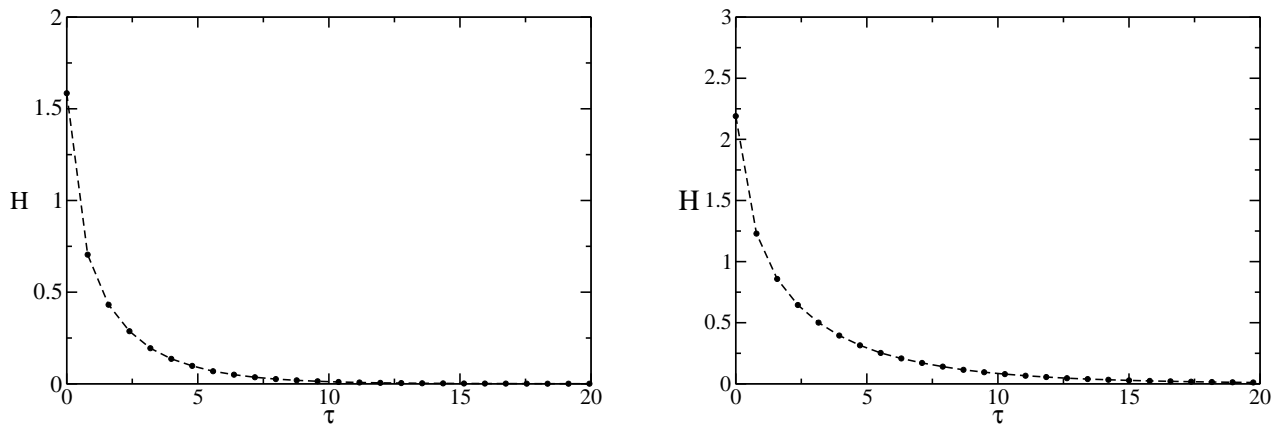

\begin{minipage}{0.49\linewidth}
\begin{center}
\includegraphics[angle=0,width=0.9\linewidth,clip]{decaimientoHplana90stoch.eps}
\end{center}
\end{minipage}
\begin{minipage}{0.49\linewidth}
\begin{center}
\includegraphics[angle=0,width=0.9\linewidth,clip]{decaimientoH95planastoch.eps}
\end{center}
\end{minipage}
\caption{ Left) DSMC results of the time evolution of ${\mathcal H}(\tau)$ for the the 
stochastic thermostat for 
$\alpha=0.9$, starting with a flat distribution. Right) The same but for 
$\alpha=0.95$. }\label{fig3}
\end{figure}

In Fig. \ref{fig3}, DSMC results are shown for a system heated by the 
stochastic thermostat. We have considered two values of the 
inelasticity, $\alpha=0.9$ and $\alpha=0.95$, with an amplitude of the noise, 
$\xi$, such that the stationary temperature is $8.80T(0)$ for $\alpha=0.9$ and 
$17.13T(0)$ for $\alpha=0.95$. In the two cases, we have started with the same 
initial flat distribution, in which all the velocities have 
the same probability in a square centered in the origin in the velocity space
\begin{displaymath}
f(\mathbf{v},0)=\left\{\begin{array}{ll}
\frac{1}{4w^2},  & \textrm{if $v_x\in[-w,w]$ and $v_y\in[-w,w]$ }\\
0,  & \textrm{otherwise}
\end{array}\right.
\end{displaymath}
with $w=\left[\frac{6T(0)}{m}\right]^{1/2}$. The results have been averaged 
over $10^5$ trajectories. Clearly, as in the 
previous case, the functional $\mathcal{H}$ decays monotonically for all times. 
Again, as in the Gaussian thermostat case, the same qualitative behavior is 
obtained for other initial conditions and values of the inelasticity.

\begin{figure}
\begin{minipage}{0.49\linewidth}
\begin{center}
\includegraphics[angle=0,width=0.95\linewidth,clip]{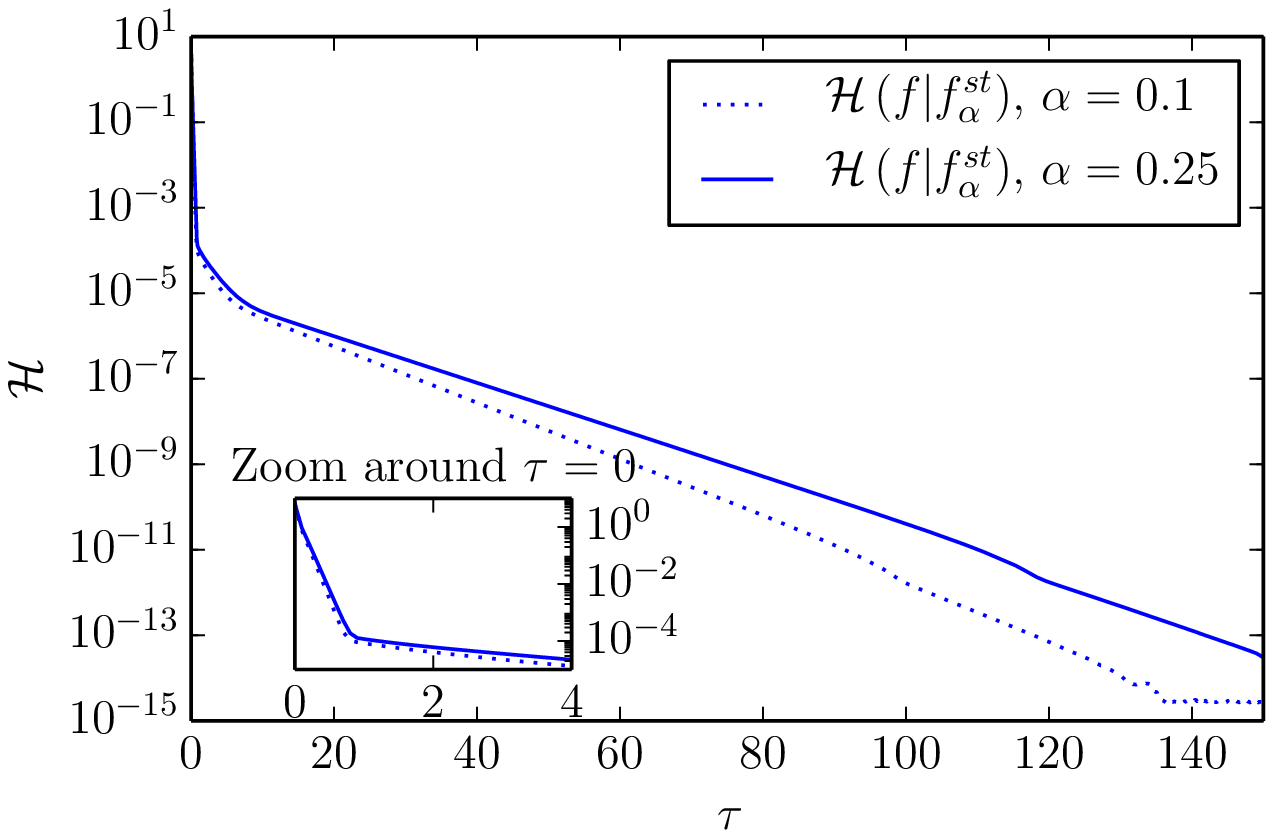}
\end{center}
\end{minipage}
\begin{minipage}{0.49\linewidth}
\begin{center}
\includegraphics[angle=0,width=0.95\linewidth,clip]{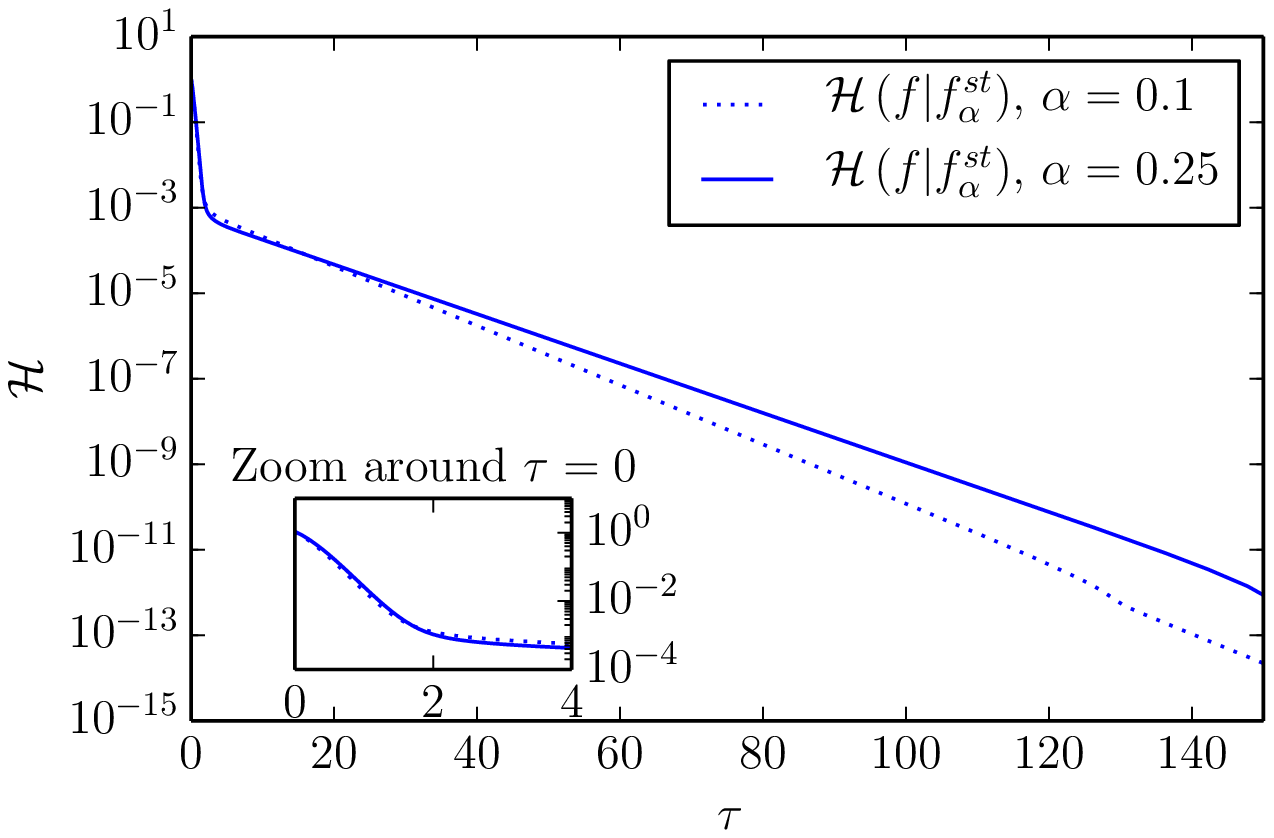}
\end{center}
\end{minipage}
\caption{ Left) Time evolution of the entropy for the nonlinear Boltzmann equation with stochastic thermostat, solved with the 
spectral scheme ($M=64$), 
for $\alpha = 0.25$ (solid line) and $\alpha = 0.1$ (dotted line), 
starting with the assymetric, three sharp peaks. 
Right)  Same, for the flat initial distribution.}\label{fig4}
\end{figure}

Finally, we present in Fig. \ref{fig4} the evolution of $\mathcal{H}$ for small normal restitution coefficients, namely $\alpha = 0.1$ (almost sticky particles) and $\alpha = 0.25$, 
in the stochastic thermostat case. 
The spectral scheme is used for these simulations.
We show our results for both the assymetric distribution \eqref{eqAssym} composed of three peaks (left) and for the flat distribution (right).
As in the other simulations, we observe in all these cases a monotone decay of the entropy functional $\mathcal H$. 
Thanks to the accuracy of the spectral scheme, and owing to its deterministic nature, we can observe this decay up to the machine precision.
Although it may be due to numerical artifacts (this behavior can be also observed in the elastic case $\alpha = 1$), this decay seems to follow two exponential regimes, a very fast one in short 
time followed by a slower one in larger time. 
Nevertheless, these decays are always exponential.

All the simulation results point in the same direction: the functional 
$\mathcal{H}$ defined by Eq. (\ref{eq:KL}) can be a good Lyapunov functional 
for the free-cooling case (Gaussian thermostat) and for the stochastic 
thermostat. It is worth emphasizing here that considering the naive
functional of the type $\int f \log f$ instead of the Kullback-Leibler distance 
\eqref{eq:KL} may lead to non-monotonic behaviour, as shown in \cite{mpv13}.

\section{Conclusions}\label{sec4}

In this paper, we have presented strong hints that the functional given by Eq. 
(\ref{eq:KL}) can play the role of a Lyapunov functional in the context of a dissipative
granular gas:  it decays monotonically in time, tending to zero
in the late time non-equilibrium steady state. These results, in agreement with those 
of Ref \cite{mpv13}, have been shown by three different 
kinds of simulation methods, for a wide class of initial conditions and a wide 
range of inelasticities, $0<\alpha<1$. Our functional --that takes the form of 
a relative entropy, or Kullback-Leibler distance-- reduces to Boltzmann's 
original $\mathcal{H}$ in the case of elastic interactions.
Its very form can be directly inspired from information theory,
where the Kullback-Leibler distance plays a prominent role \cite{CT}.

It should be noted that the asymptotic steady state $f_{st}$
enters the definition (\ref{eq:KL}), so that we cannot {\em deduce} 
the form of the non-equilibrium 
steady or scaling solution from our functional. This is at variance
with the reversible dynamics case (elastic collisions,
corresponding to $\alpha=1$ above), and is the
price for losing the energy conservation law. We emphasize that 
we have not been able to prove analytically our central result,
which requires to work at Boltzmann equation level. The situation
is simpler at $N$-body level, where a counterpart of the $H$-theorem can be shown.
We nevertheless believe our result is 
conceptually important;  monotonicity of ${\mathcal H}(t)$ with time 
is a strong statement, the derivation of which
has been the focus of some effort in the past.

In the particular free cooling case, we have shown that the
$N$-particle velocity distribution function reaches an auto-similar
regime in which all its time dependence is encoded in the
instantaneous temperature [a similar scaling has already been used in
the context of the Liouville equation (with spatial dependence)
\cite{BDS97}].  A comparable analysis can be put forward for mixtures,
binary or polydisperse, where the existence of a scaling solution
implies the coincidence of the different cooling rates. 


\appendix

\section{Evolution equation for the temperature in the 
free-cooling case}\label{a1}

Assuming the scaling form of the main text, Eq. (\ref{sHCS}), the evolution 
equation for the temperature can be straightforwardly obtained by taking the second 
velocity moment in Eq. (\ref{kacEq}). The result is 
\begin{equation}\label{eqT}
\frac{dT(t)}{dt}=-\bar{B}T(t)^{\frac{\gamma}{2}+1}, 
\end{equation}
where we have introduced the time-independent coefficient
\begin{equation}
\bar{B}=\frac{KA\langle c_{12}^{\gamma+2}\rangle}{4d}
\left(\frac{2}{m}\right)^{\gamma/2}(1-\alpha^2), 
\end{equation}
with
\begin{equation}
A=\int d\sig\hat{\sigma}_x^{\gamma+2}, \quad
\langle c_{12}^{\gamma+2}\rangle=\int d\mathbf{c}_1\int d\mathbf{c}_2c_{12}^{\gamma+2}
\varphi_2(\mathbf{c}_1,\mathbf{c}_2). 
\end{equation}
The function $\varphi_2$ is just the integrated scaled distribution, 
$\varphi_2(\mathbf{c}_1,\mathbf{c}_2)
=\int d\mathbf{c}_3\dots\int d\mathbf{c}_N\varphi_N(\mathbf{C})$. 
Eq. (\ref{eqT}) can be readily integrated, leading to the 
\begin{equation}
T(t)=\left[\frac{1}{T(0)^{\gamma/2}}
+\frac{\gamma\bar{B}}{2}t\right]^{-\frac{2}{\gamma}}, 
\qquad \textrm{for $\gamma>0$}, 
\end{equation}
and 
\begin{equation}
T(t)=T(0)e^{-\bar{B}t}\qquad  \textrm{for $\gamma=0$}. 
\end{equation}
Then, for $\gamma>0$ the temperature `forgets' the initial condition in the long 
time limit and behaves as 
\begin{equation}
T(t)\approx\left(\frac{\gamma\bar{B}}{2}t\right)^{-2/\gamma}, 
\qquad\textrm{for $t \to \infty$}. 
\end{equation}

We point now that the scaling given by Eq. (\ref{sHCS}) is
compatible with Kac's equation, Eq. (\ref{kacEq}). By substituting 
Eq. (\ref{sHCS}) into Eq. (\ref{kacEq}), the following equation for $\varphi_N$ 
is obtained
\begin{equation}\label{EqVarphi}
\frac{1}{Kv_H^{\gamma+1}}\frac{dv_H(t)}{dt}
\frac{\partial}{\partial\mathbf{C}}\cdot\mathbf{C}
\varphi_N(\mathbf{C})+\frac{1}{N}\sum_{i<j}L(\mathbf{c}_i,\mathbf{c}_j)
\varphi_N(\mathbf{C})=0, 
\end{equation}
where $\mathbf{C}=\mathbf{V}/v_H(t)$. As a consequence of Eq. (\ref{eqT}), 
the term $\frac{1}{Kv_H^{\gamma+1}}\frac{dv_H(t)}{dt}$ is time-independent, and 
thus, Eq. (\ref{EqVarphi}) is consistent. 

\section{Evaluation of $\frac{d\mathcal{H}_N(s)}{ds}$}
\label{a2}

Before evaluating $\frac{d\mathcal{H}_N(s)}{ds}$, let us note that Eq. 
(\ref{kacEqC}) can be written as a Master equation
\begin{equation}
\frac{\partial}{\partial s}\phi_N(\mathbf{C},s)=\int d\mathbf{U}
\mathcal{W}(\mathbf{C}|\mathbf{U})\phi_N(\mathbf{U},s)-
\int d\mathbf{U}
\mathcal{W}(\mathbf{U}|\mathbf{C})\phi_N(\mathbf{C},s), 
\end{equation}
with the transition probabilities per unit time, 
$\mathcal{W}(\mathbf{U}|\mathbf{C})$, given by
\begin{equation}
\mathcal{W}(\mathbf{U}|\mathbf{C})=\mathcal{W}_C(\mathbf{U}|\mathbf{C})+
\mathcal{W}_T(\mathbf{U}|\mathbf{C}), 
\end{equation}
where
\begin{equation}
\mathcal{W}_C(\mathbf{U}|\mathbf{C})=\frac{1}{N}\sum_{i<j}\int d\sig
(\mathbf{c}_{ij}\cdot\sig)^\gamma\delta(\mathbf{u}_i-\mathbf{c}_i')
\delta(\mathbf{u}_j-\mathbf{c}_j')\Pi_{k\ne i,j}\delta(\mathbf{u}_k-\mathbf{c}_k), 
\end{equation}
is the contribution due to collision and 
\begin{equation}
\mathcal{W}_T(\mathbf{U}|\mathbf{C})=-\frac{B}{2}\mathbf{C}\cdot
\frac{\partial}{\partial\mathbf{U}}\delta(\mathbf{U}-\mathbf{C}), 
\end{equation}
the contribution of the thermostat. 

Following the same steps as van Kampen \cite{kampen}, it is obtained that
\begin{eqnarray}\label{ap1}
\frac{d\mathcal{H}_N(s)}{ds}=\int d\mathbf{C}\int d\mathbf{U}
\mathcal{W}(\mathbf{C}|\mathbf{U})\phi_N^{st}(\mathbf{U})
\left\{[x(\mathbf{U},s)-x(\mathbf{C},s)]h'[x(\mathbf{C},s)]\right. \nonumber\\
\left.
+h[x(\mathbf{C},s)]
-h[x(\mathbf{U},s)]\right\}, 
\end{eqnarray}
where 
\begin{equation}
x(\mathbf{C},s)=\frac{\phi_N(\mathbf{C},s)}{\phi_N^{st}(\mathbf{C})}. 
\end{equation}
It can be seen that the contribution coming from the thermostat 
vanishes. Indeed, 
\begin{eqnarray}\label{Tcont}
\int d\mathbf{C}\int d\mathbf{U}
\mathcal{W}_T(\mathbf{C}|\mathbf{U})\phi_N^{st}(\mathbf{U})
\left\{[x(\mathbf{U},s)-x(\mathbf{C},s)]h'[x(\mathbf{C},s)]
+h[x(\mathbf{C},s)]
-h[x(\mathbf{U},s)]\right\}\nonumber\\
=\frac{B}{2}\int d\mathbf{C}\int d\mathbf{U}\delta(\mathbf{U}-\mathbf{C})
\phi_N^{st}(\mathbf{U})\mathbf{U}\cdot\frac{\partial}{\partial\mathbf{C}}
\left\{[x(\mathbf{U},s)-x(\mathbf{C},s)]h'[x(\mathbf{C},s)]\right.\nonumber\\
\left.+h[x(\mathbf{C},s)]
-h[x(\mathbf{U},s)]\right\}=0. 
\end{eqnarray}
The collisional counterpart can be written as
\begin{eqnarray}\label{Ccont}
\int d\mathbf{C}\int d\mathbf{U}
\mathcal{W}_C(\mathbf{C}|\mathbf{U})\phi_N^{st}(\mathbf{U})
\left\{[x(\mathbf{U},s)-x(\mathbf{C},s)]h'[x(\mathbf{C},s)]
  +h[x(\mathbf{C},s)]-h[x(\mathbf{U},s)]\right\}\nonumber\\
=\frac{N-1}{2}\int d\mathbf{C}
\int d\sig (\mathbf{c}_{12}\cdot\sig)^\gamma\phi_N^{st}(\mathbf{C})
\left\{h[x(\mathbf{C}',s)]-h[x(\mathbf{C},s)]+h'[x(\mathbf{C}',s)]
[x(\mathbf{C},s)-x(\mathbf{C}',s)]\right\}, \nonumber\\
\end{eqnarray}
where we have introduced the notation 
$\mathbf{C}'=\{\mathbf{c}_1',\mathbf{c}_2',\mathbf{c}_3,\dots,\mathbf{c}_N\}$ 
and we have used that, due to the invariance of $\phi_N(\mathbf{C},s)$ under 
the change of labels $\mathbf{c}_i\leftrightarrow\mathbf{c}_j$ for any $i$ and 
$j$, the $N(N-1)/2$ collisional terms are equal. By substituting Eqs 
(\ref{Tcont}) and (\ref{Ccont}) into (\ref{ap1}), the expression of the main 
text is obtained. 

\section{Consistency of Kac's equation in the stochastic 
thermostat case}\label{a3}

In this appendix, it is shown that the conservation of the total momentum 
assumed by the thermostat is essential to ensure that Eq. (\ref{kacEqTS}) be 
compatible with the existence of a stationary state. 
We assume that a stationary solution of Eq. (\ref{kacEqTS}) exists, 
$\rho_N^{st}(\mathbf{V})$. It then fulfills
\begin{equation}\label{STss}
\frac{K}{N}\sum_{i<j}L(\mathbf{v}_i,\mathbf{v}_j)\rho_N^{st}(\mathbf{V})
+\xi_0^2T(\mathbf{V})\rho_N^{st}(\mathbf{V})=0. 
\end{equation}
Let us multiply the equation by $v_1^2$ and integrate. The collisional 
contribution is 
\begin{eqnarray}
\frac{K}{N}\sum_{i<j}\int \, {\rm d}\mathbf{v}v_1^2L(\mathbf{v}_i,\mathbf{v}_j)
\rho_N^{st}(\mathbf{V})\nonumber\\
=\frac{K(N-1)}{2N}\int \, {\rm d}\mathbf{v}\rho_N^{st}(\mathbf{V})
\int d\sig (\sig\cdot\mathbf{v}_{12})^\gamma(b_{\sig}-1)(v_1^2+v_2^2)\nonumber\\
=-\frac{K(N-1)}{4N}(1-\alpha^2)\int \, {\rm d}\mathbf{v}\rho_N^{st}(\mathbf{V})
\int d\sig (\sig\cdot\mathbf{v}_{12})^{\gamma+2}, 
\end{eqnarray}
where use has been made of
\begin{equation}\label{inEn}
(b_{\sig}-1)(v_1^2+v_2^2)=-\frac{1-\alpha^2}{2}(\sig\cdot\mathbf{v}_{12})^2. 
\end{equation}
The thermostat contribution is
\begin{equation}
\xi_0^2\int \, {\rm d}\mathbf{v}v_1^2T(\mathbf{V})\rho_N^{st}(\mathbf{V})
=\frac{\xi_0^2}{2}\int \, {\rm d}\mathbf{v}\rho_N^{st}(\mathbf{V})
\frac{\partial^2}{\partial v_1^2}v_1^2=d\xi_0^2, 
\end{equation}
where the cross-terms do not contribute. The obtained equation is 
\begin{equation}\label{obEq}
\frac{K(N-1)}{4N}(1-\alpha^2)\int \, {\rm d}\mathbf{v}\rho_N^{st}(\mathbf{V})
\int d\sig (\sig\cdot\mathbf{v}_{12})^{\gamma+2}=d\xi_0^2. 
\end{equation}
Keeping in mind the above `steady-state constraint', 
let us now multiply Eq. (\ref{STss}) by $\mathbf{v}_1\cdot\mathbf{v}_2$ and 
integrate. The collisional contribution is 
\begin{eqnarray}
\frac{K}{N}\sum_{i<j}\int \, {\rm d}\mathbf{v}\mathbf{v}_1\cdot\mathbf{v}_2
L(\mathbf{v}_i,\mathbf{v}_j)\rho_N^{st}(\mathbf{V})\nonumber\\
=\frac{K}{N}\int \, {\rm d}\mathbf{v}\rho_N^{st}(\mathbf{V})
\int d\sig (\sig\cdot\mathbf{v}_{12})^\gamma
(b_{\sig}-1)\mathbf{v}_1\cdot\mathbf{v}_2\nonumber\\
=\frac{K}{4N}(1-\alpha^2)\int \, {\rm d}\mathbf{v}\rho_N^{st}(\mathbf{V})
\int d\sig (\sig\cdot\mathbf{v}_{12})^{\gamma+2}, 
\end{eqnarray}
where it has been used that
\begin{equation}\label{inSP}
(b_{\sig}-1)\mathbf{v}_1\cdot\mathbf{v}_2
=\frac{1-\alpha^2}{4}(\sig\cdot\mathbf{v}_{12})^2. 
\end{equation}
From the conservation of total momentum in a collision, we further have 
$(b_{\sig}-1)(\mathbf{v}_1+\mathbf{v}_2)^2=0$, so that
\begin{equation}
(b_{\sig}-1)\mathbf{v}_1\cdot\mathbf{v}_2=-\frac{1}{2}
(b_{\sig}-1)(v_1^2+v_2^2), 
\end{equation}
from Eqs. (\ref{inEn}) and (\ref{inSP}). 
The thermostat contribution is
\begin{eqnarray}
\xi_0^2\int \, {\rm d}\mathbf{v}\mathbf{v}_1\cdot\mathbf{v}_2T(\mathbf{V})
\rho_N^{st}(\mathbf{V})\nonumber\\
=-\frac{\xi_0^2}{N-1}\int \, {\rm d}\mathbf{v}\rho_N^{st}(\mathbf{V})
\frac{\partial}{\partial\mathbf{v}_1}\frac{\partial}{\partial\mathbf{v}_2}
\mathbf{v}_1\cdot\mathbf{v}_2=-\frac{d\xi_0^2}{N-1},  
\end{eqnarray}
where only the cross-terms contribute. The obtained equation is equivalent to 
Eq. (\ref{obEq}) but, as indicated above, only because of the presence of the 
cross terms [second term on the right hand side of Eq. \eqref{coefB}].
In other words, it is essential that the thermostat conserve total 
momentum, otherwise, the center-of-mass of the system undergoes a
Brownian motion in velocity space, incompatible with the existence 
of a steady state \cite{PSD13}.


\begin{thebibliography}{10}
\bibitem{rque1}
It should be noted here that Boltzmann's functional clearly is a
forerunner of Shannon's information measure, introduced more than 70 years 
later \cite{Shannon}. This measure is of paramount 
importance in data and signal analysis or transmission.
It should also be stressed that the $H$-theorem today still is
a remarkable source of inspiration to attack many mathematical
problems (see e.g. Ref \cite{Villani}), socio-physics questions (see e.g. Ref \cite{Apenko}) 
or to find new classes of time dependent 
exact solutions to the Boltzmann equation under confinement \cite{dgo}.
\bibitem{rque55}
Molecular Dynamics simulations are more demanding in the driven case,
particularly so if small densities are considered.
\end{thebibliography}
\end{document}